\documentclass[twocolumn,prl,aps,showkeys,showpacs]{revtex4-1}

\usepackage{ucs}
\usepackage[english]{babel}
\usepackage{amsmath}
\usepackage{amssymb,amsfonts}
\usepackage{color}

\usepackage{graphicx}
\usepackage{dcolumn}
\usepackage{bm}

\usepackage{bbold}

\usepackage{subfigure}

\usepackage{tikz}
\usetikzlibrary{shapes.misc}

\usepackage{mathrsfs}

\usepackage{scrpage2}
\usepackage{hyperref}

\newcommand{\dBv}{\Delta B}

\usepackage[normalem]{ulem}

\begin{document}

\title{Stabilizing a discrete time crystal against dissipation}
\author{Leon Droenner}%
\affiliation{Technische Universit\"at Berlin, Institut f\"ur Theoretische Physik, Nichtlineare Optik und Quantenelektronik, Hardenbergstra{\ss}e 36, 10623 Berlin, Germany}
\author{Regina Finsterh\"olzl}
\affiliation{Technische Universit\"at Berlin, Institut f\"ur Theoretische Physik, Nichtlineare Optik und Quantenelektronik, Hardenbergstra{\ss}e 36, 10623 Berlin, Germany}
\author{Markus Heyl}
\affiliation{Max-Planck-Institut f\"ur Physik komplexer Systeme, N\"othnitzer Straße 38, 01187 Dresden, Germany}
\author{Alexander Carmele}
\email{alex@itp.tu-berlin.de}
\affiliation{Technische Universit\"at Berlin, Institut f\"ur Theoretische Physik, Nichtlineare Optik und Quantenelektronik, Hardenbergstra{\ss}e 36, 10623 Berlin, Germany}


\begin{abstract}
Eigenstate phases such as the discrete time crystal exhibit an inherent instability upon the coupling to an environment, which restores equipartition of energy and therefore acts against the protecting nonergodicity. Here, we demonstrate that a discrete time crystal can be stabilized against dissipation using coherent feedback. For a kicked random Ising chain subject to a radiative decay, we show that the time crystalline signal can survive through a mechanism of constructive interference upon reflecting the emitted photons by a mirror. We introduce a matrix product operator algorithm to solve the resulting non-Markovian dynamics. We find that the stabilization mechanism is robust against weak imperfections.
\end{abstract}

\maketitle
\emph{Introduction.}\textbf{--}
Nonergodicity provides a mechanism to generate phase structures of quantum matter inaccessible within the thermodynamic paradigm by avoiding equipartition of energy through selectively occupying eigenstates~\cite{2013PhRvB..88a4206H,nandkishore2015many}.
Many-body localized (MBL) spin glasses~\cite{2013PhRvB..88a4206H,2014PhRvL.113j7204K} or discrete time crystals (DTC)~\cite{khemani2016phase,else2016floquet} constitute prime examples of such resulting eigenstate phases.
The protecting nonergodicity in interacting nonintegrable systems can be induced by imposing strong quenched disorder~\cite{Basko_MBL,nandkishore2015many,2015ARCMP...6..383A} or by dynamical constraints that can lead to disorder-free localization in gauge theories~\cite{2017PhRvL.118z6601S,2017PhRvL.119q6601S,2018PhRvL.120c0601B} or quantum many-body scars~\cite{2018NatPh..14..745T,ho2019periodic}.
Today, it is possible to realize the necessary unitary nonergodic evolution also experimentally in so-called quantum simulators, which has led to the observation of MBL~\cite{Schreiber2015,Smith2016,choi2016exploring} and DTCs~\cite{Zhang2017,Choi2017} for instance.
However, dissipation represents a major challenge in this context since an even weak coupling to an environment generically restores equipartition implying an inherent instability of eigenstate phases against a coupling to an environment~\cite{2007PhRvB..76e2203B,PhysRevLett.116.237203,Coupled_to_a_bath_lindblad_MBL,dephasing_MBL_znidaric,2018PhRvL.121z7603L,2017PhRvB..95s5135L,droenner2017boundary}.

\begin{figure}
	\center
	\includegraphics[width=\columnwidth]{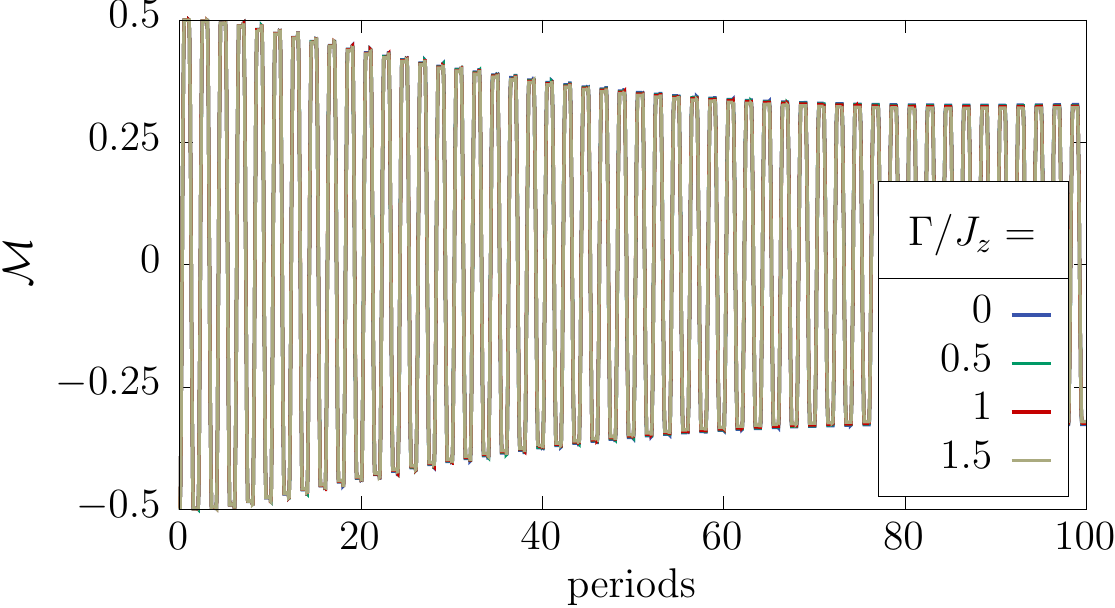}
	\caption{Dynamics of the staggered magnetization $\mathcal{M}$ for a discrete time crystal in a kicked random transverse-field Ising chain of $N=40$ spins coupled to an environment with coherent feedback. Even for a significant coupling to the environment $\Gamma/J_z$ the time-crystalline signal survives and is only hardly distinguishable compared to the case without environment $\Gamma=0$. For this data we have used a parameter regime of imperfect kicking with $\epsilon/J^z=0.15$ and $h^x/J^z = J^x/J^z=0.1$.}
	\label{fig:1}
\end{figure}

In this work, we show that it is possible to stabilize DTCs against dissipation utilizing coherent feedback.
Specifically, we study a paradigmatic model for a DTC realized in a MBL Ising chain of spin-1/2 degrees of freedom subject to a radiative decay, which 
leads to a collapse of the time-crystalline order~\cite{2017PhRvB..95s5135L}.
Upon reflecting the emitted photons using a mirror, we find regimes of constructive interference when varying the distance to the mirror~\cite{PhysRevLett.116.093601,carmele2013,nemet2016enhanced,grimsmo2015time,zhang2017quantum}. 
Using such non-Markovian dynamics, the time-crystalline signal can be stabilized against coupling to the environment, as we show in Fig.~\ref{fig:1}.
Importantly, we find that the stabilization is robust for deviations from the optimal constructive distance.
For the purpose of studying the dynamics in the anticipated system we develop a matrix product operator technique allowing us to treat systems of up to $N=40$ spins including
$40$ individual reservoirs with non-Markovian dynamics.
Such a large system size becomes accessible due to the fact that quantum feedback 
suppresses entanglement growth in the system-reservoir dynamics while MBL prevents such a growth within the many-body system itself.
This method captures the non-Markovian character of the environment, which is crucial for the stabilization via constructive interference.
\emph{Model.}\textbf{--}
We study the stabilization of a DTC via coherent feedback for a paradigmatic model.
Specifically, we consider a Floquet system with a periodically time-dependent Hamiltonian $\mathcal{H}(t+T)=\mathcal{H}(t)$ realizing the following time evolution operator $U$ over one period $T$ ($\hbar=1$):
\begin{align}
U=e^{-iT \mathcal{H}_I/2}e^{-iT\mathcal{H}_T/2},
\end{align}
with two piecewise constant Hamiltonians over each half-period $T/2$ where
\begin{align}
\mathcal{H}_T&=\sum_{i=1}^N (\Omega-\epsilon) \sigma_i^x~, \label{eq:floquet-drive}\\
\mathcal{H}_I&=\sum_{i=1}^{N-1}J^z_i\sigma_i^z\sigma_{i+1}^z+\sum_{i=1}^{N-1}J^x_i\sigma_i^x\sigma_{i+1}^x +\sum_{i=1}^N h^x_i\sigma^x_i~.\label{IsingMBL-SG}
\end{align}
Here, we fix $\Omega=\pi/T$, such that for $\epsilon=0$ the impact of $\mathcal{H}_T$ is to flip each spin of the chain.
A nonzero $\epsilon$ constitutes a perturbation from the perfect polarization flip, which we include to study the robustness of the DTC.
During the second half period the dynamics is driven by a random transverse-field Ising chain including a weak integrability-breaking transverse coupling.
We choose all couplings from uniform distributions, i.e.,  $J^z_i\in\left[-J^z,J^z\right]$, $J^x_i\in\left[-J^x,J^x\right]$, and $h^x_i\in\left[-h^x,h^x\right]$.
For the remainder of this work, we consider a limit, where the static $\mathcal{H}_I$ realizes an MBL spin-glass and therefore ensures the spatial ordering necessary for a DTC.
Concretely, we take $h^x/J^z=J^x/J^z=0.1$ for our simulations throughout this work.
For numerical convenience, as will be discussed below, we furthermore choose $J^z T = 0.05$.
We couple our spin chain to external bosonic mode continua.
We consider two cases, which, as we will find, lead to the equivalent dynamics in our case.
This includes to couple each spin in the chain to its own bath with a Hamiltonian:
\begin{align}
H_{\mathcal{D}}&=\sum_{i=1}^N\int d\omega \, \omega b_i^\dagger(\omega)b_i(\omega)\notag\\
&+ \sum_{i=1}^N\int d\omega \, \left[G_i(\omega) b_i^\dagger(\omega) \sigma_i^-  + h.c.\right]~,
\label{eq:Hfb}
\end{align} 
where $b_i(\omega)$ annihilates a boson of energy $\omega$ at lattice site $i$.
For a global bath, we consider only a single species of bosons $b(\omega)$ and a site-independent $G_i(\omega) = G(\omega)$.
In case the coupling $G_i(\omega)$ is sufficiently weak and frequency-independent, this model describes spontaneous radiative decay of the two-level systems upon emitting photons into the continuum, which can be accurately modeled via Lindblad dissipators.
Such a coupling to an environment has been shown to lead to a decay of the time-crystalline signal~\cite{2017PhRvB..95s5135L}.
It is the main purpose of this work to go beyond such a Lindblad master equation treatment and to take into account the effect of a non-Markovian environment.
Specifically, we aim to consider a boundary condition for the mode continuum such as a distant mirror reflecting the photons back onto the spin system~\cite{dorner02,PhysRevA.93.053807, doi:10.1080/09500340.2017.1363919}.
\begin{figure*}
\centering
\flushleft{(a)\hspace{7cm}(b)}\newline
\includegraphics[width=0.37\textwidth]{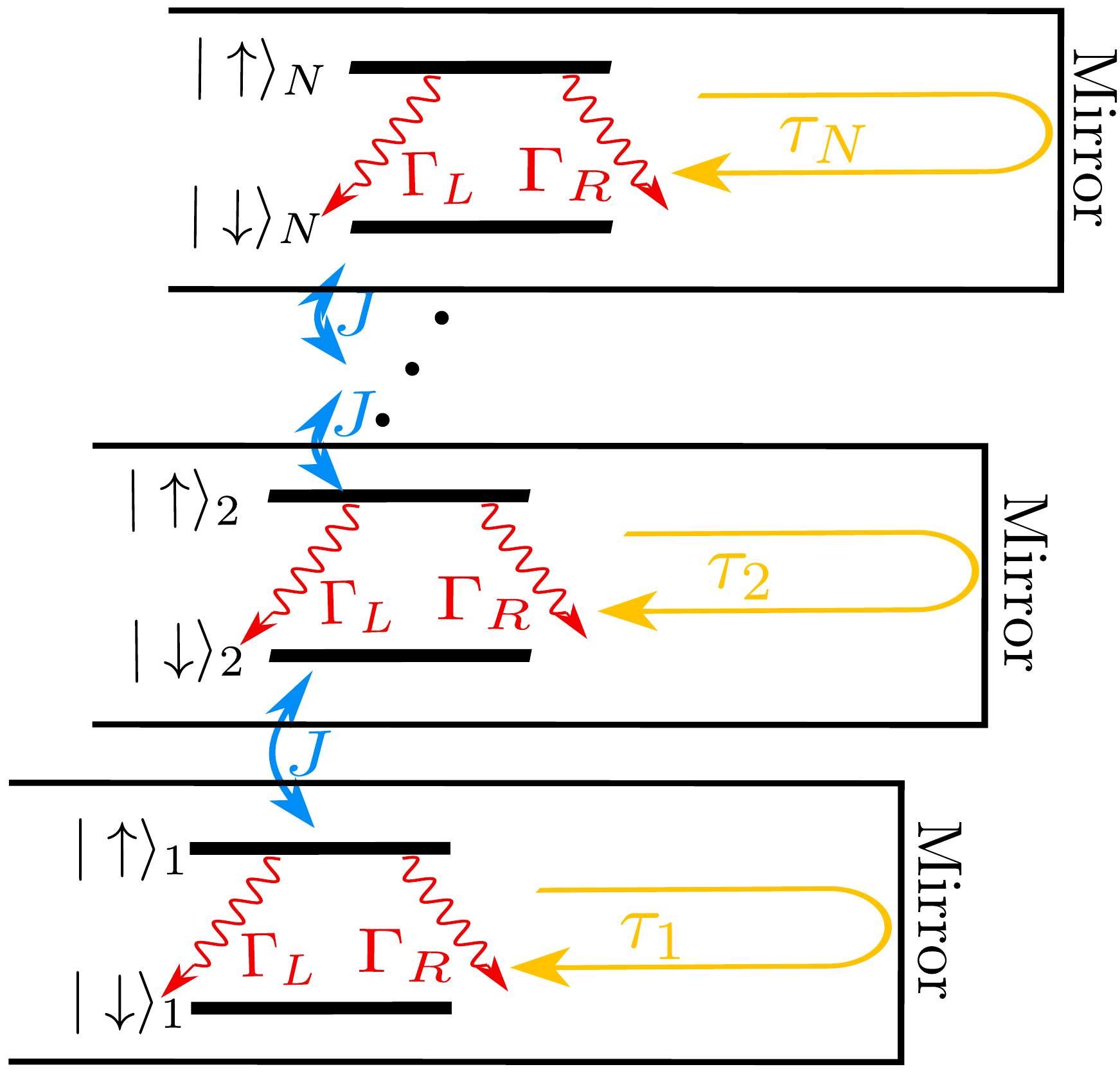}
\includegraphics[width=0.57\textwidth]{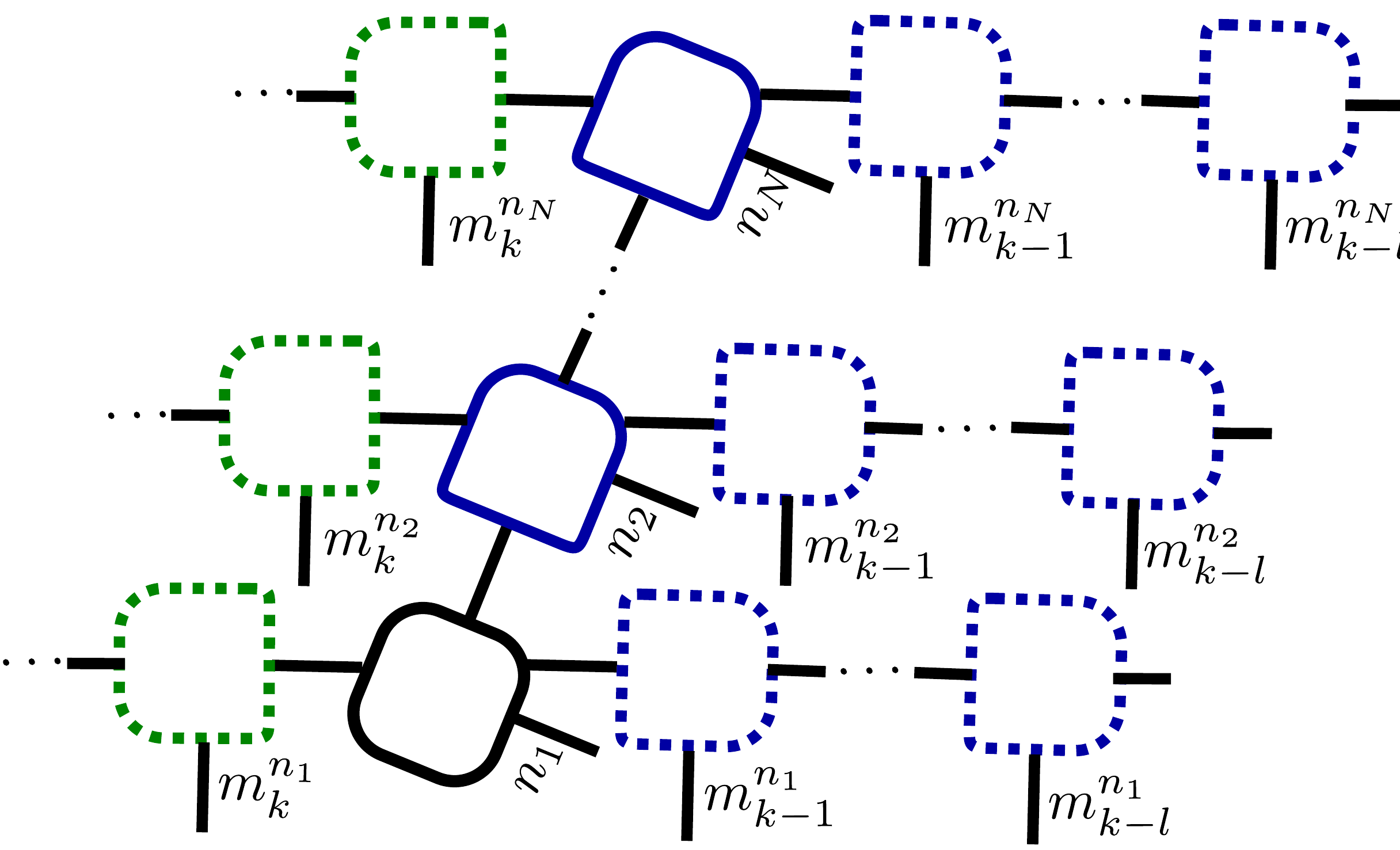}
\caption{(a): Model of the spin-chain coupled to external structured reservoirs. Each spin is subject to a decay into its individual environment.  The spins are coupled to each other with Ising interaction $J$. Each individual reservoir contains a boundary condition which is sketched as a semi-infinite waveguide, where excitations are reflected and interact again with the system after a time-delay $\tau_{n_i}$. (b): The MPS is two-dimensional, where the dashed MPS with physical indices $n_i$ (solid boxes) corresponds to the many-body system. The horizontal MPSs with physical indices $m^{n_i}_k$ (dashed boxes) correspond to the reservoirs expressed in the time-bin basis. The color denotes the orthogonality (green - left orthogonal, blue - right orthogonal and black is the orthogonality center).}
\label{fig:model_DDTC}
\end{figure*} 
Alternatively, we can imagine that each spin is put inside a semi-infinite waveguide, which is sketched in \autoref{fig:model_DDTC} (a), where the coupling has a sinusoidal frequency dependency \cite{PhysRevLett.116.093601, droenner2018two}
\begin{equation}
	G_{i}(\omega)=i\left(\sqrt{\frac{\Gamma_R}{2\pi}}e^{-i\omega\tau_{n_i}/2}-\sqrt{\frac{\Gamma_L}{2\pi}}e^{i\omega\tau_{n_i}/2}\right) \, .
\end{equation}
The parameters $\Gamma_L$ and $\Gamma_R$ denote the couplings to the left infinite and the right closed side, respectively.
Due to the structured reservoir, excitations return and interact again with the system after a time-delay $\tau_{n_i}=2L_{ni}/c$, where $L_{n_i}$ is the distance to the closed end and $c$ the speed of light in the waveguide.
In our chosen model, the interaction to the environment takes place during the whole period $T$, which includes the transverse field and the interaction.
Let us emphasize, however, that our main results are not just limited to this case.

We prepare our system initially in a N\'eel state $|\psi(0)\rangle=|\uparrow\downarrow\uparrow \downarrow\dots\rangle$ and detect the time-crystalline order via the staggered magnetization
\begin{align}
{\mathcal M}= \frac{1}{N}\sum_{i=1}^N (-1)^{i+1}  \sigma^z_i~.\label{eq:stagg_magnetization}
\end{align}
For the purely unitary dynamics without coupling to an environment, this scenario is equivalent to the initial fully polarized state and measuring the magnetization, as is mostly studied in the context of DTCs, through a unitary transformation flipping every second spin.
While highly polarized states can be close to dark states of the dynamics, the N\'eel state cannot.
In this way we make sure that the observed stabilization is solely due to the coherent feedback and not due to intermediate dark states.

\emph{Method.}\textbf{--} As discussed before, we consider two cases: A global reservoir and an individual reservoir for each spin.
As we will show, the resulting dynamics is, however, very similar. To deal with a non-Markovian environment, we compute the global reservoir similar to the Ref. \cite{PhysRevLett.116.093601,droenner2018two} with the quantum stochastic Schr\"odinger equation (QSSE), where we store all states of the many-body system inside a single tensor. 
For this reason,  assuming a global reservoir, we are limited to small system sizes. For an efficient computation, we choose an individual reservoir for each spin. This allows us to formulate the QSSE method to compute non-Markovian system-reservoir dynamics in terms of a matrix product operator (MPO) to deal with the many-body system. With this we are able to explore much larger system sizes when the entanglement to reservoir degrees of freedom is suppressed via constructive feedback ($\phi_{n_i}=\pi$), which is the case when the DTC is stabilized.\\
Using a time-bin basis $|m_k^{n_i}\rangle$ for the QSSE picture, each spin $n_i$ has its own reservoir expressed as an MPS where the physical index $m_k^{n_i}$ corresponds to the time-bin at time interval $\Delta t= t_{k+1}-t_k$, where $\Delta t$ is the numerical time-step. The time increment time-evolution operator for the dissipation part then reads
\begin{align}
U_\mathcal{D}(t_{k+1},t_{k})=&\prod_{i=1}^N\exp\Big[-\left( \sqrt{\Gamma_R} \dBv_{n_i}(t_{k-l}) e^{-i\phi_{n_i}}\right.\notag\\
&\left.+\sqrt{\Gamma_L} \dBv_{n_i}(t_k) \right) \sigma_i^{+}  +h.c.\Big]~,
 \label{eq:time-evolution-DTC-dissipation}
\end{align}
with feedback phase $\phi_{n_i}$ and quantum noise operators $\Delta B_{n_i}^{(\dagger)}(t_k)$ which describe the annihilation (creation) of reservoir excitations in the state $|m_k^{n_i}\rangle$. 
In contrast to conventional MPS algorithms, our MPS is two dimensional, which is shown in \autoref{fig:model_DDTC} (b). The MPS of the many-body system is sloped with solid boxed. Each tensor connects to the respective horizontal reservoir MPS, where the dashed boxes represent the respective time-bins. We evolve the two-dimensional MPS in time in applying the respective time-evolution operator during the Floquet period which are formulated as MPOs: The time-evolution operator for the transverse field consists of $N$ local operations, where a formulation as an MPO is straightforward. The time-evolution operator for the interacting part is expressed as an MPO in performing a Suzuki-Trotter approximation, involving an error which scales as $\mathcal{O}(\Delta t^2)$. However, to include the dissipation, some remarks about the action of the time-evolution operator in Eq. \ref{eq:time-evolution-DTC-dissipation} on the two-dimensional MPS are in order: $U_{\mathcal{D}}$ is local in terms of the many-body system. However, it also acts on two different times, the future time-bin $m_k^{n_i}$ via operator $\dBv_{n_i}(t_k)$ and on the past time-bin $m_{k-l}^{n_i}$ via $\dBv_{n_i}(t_{k-l})$ which is called the feedback time-bin with $t_{k-l}=k\Delta t-\tau$. Thus, each local tensor of the time-evolution operator acts on three tensors: The system tensor, corresponding to the physical index $n_i$, the future time-bin $m_k^{n_i}$ which is next to the system tensor and on the feedback time-bin $m_{k-l}^{n_i}$. Note that the feedback time-bin is a distant tensor, where the distance depends on $\Delta t/\tau$. To deal with this long-range interactions, we apply swap operations \cite{PhysRevLett.116.093601} on each reservoir MPS to bring all feedback time-bins next to the respective system-bin.
With the assumption of individual reservoirs for each spin, the dissipative time-evolution operator in Eq. \ref{eq:time-evolution-DTC-dissipation} can be multiplied into both system time-evolution operators respectively, without destroying the MPO form. 
Concretely, we will choose the following parameters for the numerical simulations shown throughout this Letter.
\begin{figure}
	\center
	\includegraphics[width=\columnwidth]{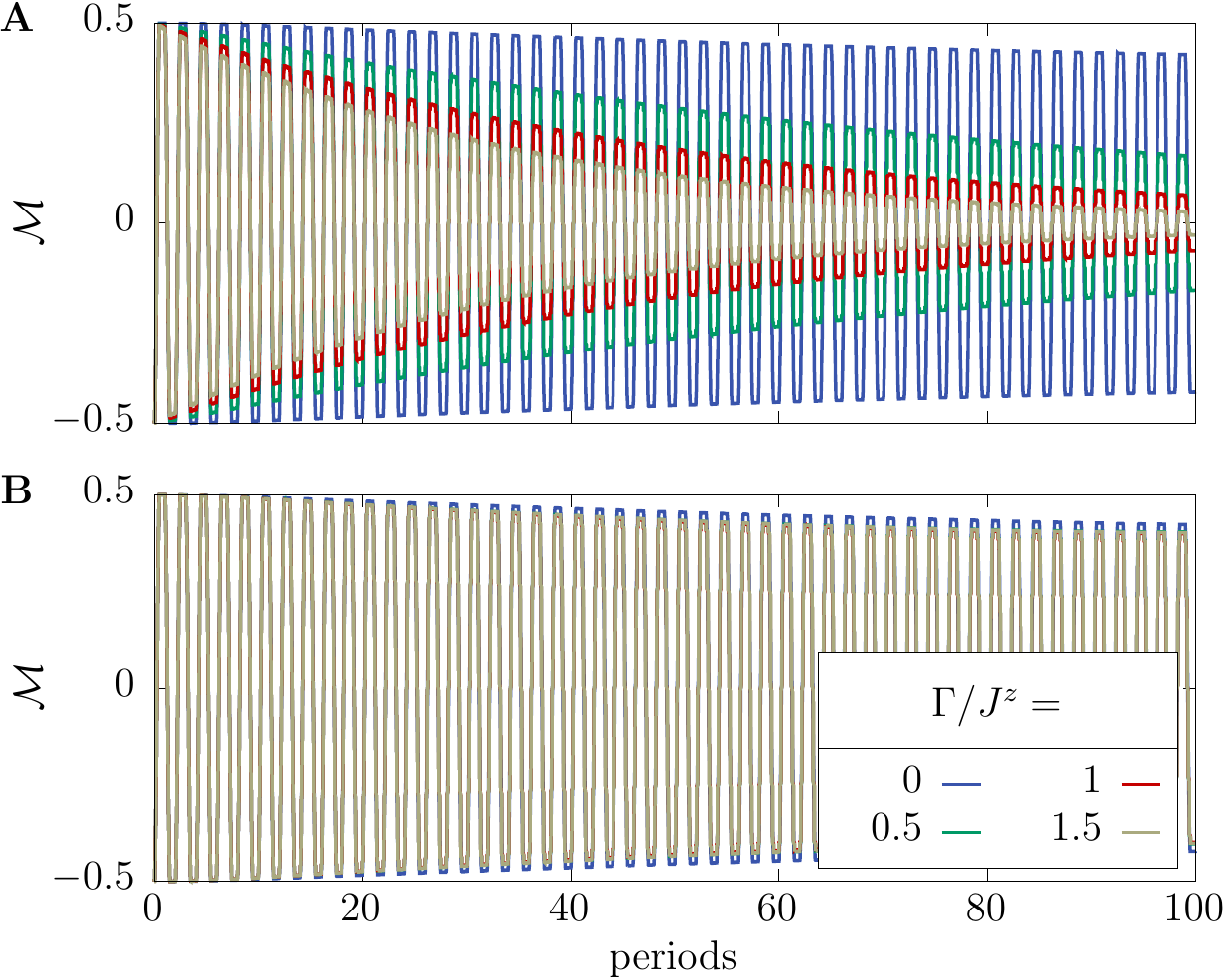}
	\caption{Time trace of the staggered magnetization comparing the evolution of the system subject to Markovian bath (A) to utilizing coherent feedback (B). For Markovian dynamics (A), the DTC decays with a larger rate upon  increasing the reservoir coupling $\Gamma$. For non-Markovian dynamics (B), the DTC signal becomes almost independent of the coupling strength $\Gamma$ leading to a stabilization of the DTC.}
	\label{fig:time_trace}
\end{figure}
As quantum feedback we consider a semi-infinite waveguide, where the couplings to both sides are equal $\Gamma_L = \Gamma_R= \Gamma$.
In order to achieve the stabilization of the DTC, the feedback phase $\phi_{n_i}=\omega_s\tau_{n_i}$ for each spin on the lattice is the crucial parameter, with $\omega_s$ being the frequency difference between spin-up and spin-down. 
We find that the feedback acts constructively on the spin-dynamics, yielding stabilization, when all $\phi_{n_i}=(2k-1)\pi$, with integer number $k$.
Since $\omega_s$ is identical for all the involved lattice sites, we can assume a site-independent feedback phase $\tau = \tau_{n_i}$ for all $i$.
In the following, we will work in the limit of fast feedback $J_z\tau = 2 \times 10^{-4}$.
Thus, the reflected photon returns quickly, which is justified also from an experimental perspective, since photons can be assumed to travel fast compared to any internal dynamics of the spin system. 

Practically, the phases $\phi_{n_i}$ can be tuned by varying the time delay of the reflected photon, which is set by the distance covered by traveling photon and therefore by the distance of the mirror.
We note that when $\phi_{n_i}=2k\pi$, the opposite is the case and feedback dynamics result in an even faster equipartitioning than for the purely Markovian dissipation. 
As we show later, the stabilization does not require absolute fine-tuning of the feedback phase $\phi_{n_i}$.
Deviations from the case of optimal constructive interference lead to a slight decrease of the time-crystalline signal.
The stabilization, however, still remains effective.

\emph{Feedback stabilization.}\textbf{--}
After having outlined the numerical method to solve for the dynamics, we now aim to discuss our main findings, which we show in Fig.~\ref{fig:time_trace}.
In the presence of a Markovian environment, without the coherent feedback, the time-crystalline signal decays on a time scale inversely proportional to the coupling strength to the environment, see Fig.~\ref{fig:time_trace}(A).
Here, we also employ the QSSE formalism by simulating the dynamics with a coupling only to the unstructured part of the reservoir, i.e. by setting $\Gamma_R=0$.
However, for the Markovian case this method is highly inefficient due to the fast
entanglement growth between system and reservoir states. 
Consequently, the system approaches equipartition leading to a collapse of the signal.
Importantly, as we show in Fig.~\ref{fig:time_trace}(B), the time crystal can be stabilized significantly upon adding coherent feedback, i.e. a memory kernel to the reservoir.
In particular and surprisingly, we find that the resulting dynamics is nearly independent on the coupling to the non-Markovian environment $\Gamma$ indicating that the feedback is capable to compensate the influence of the environment by suppressing the aforementioned entanglement growth between system and reservoir states.
Note, in Fig.~\ref{fig:time_trace}(A) and (B) we used a small system size due to the 
extreme fast entanglement growth for the Markovian case to compare non-Markovian and Markovian dynamics on equal footing. 
Furthermore, we have chosen a global reservoir for all spins to reduce the number of states in the MPS algorithm.
\begin{figure}
\center
\includegraphics[width=\columnwidth]{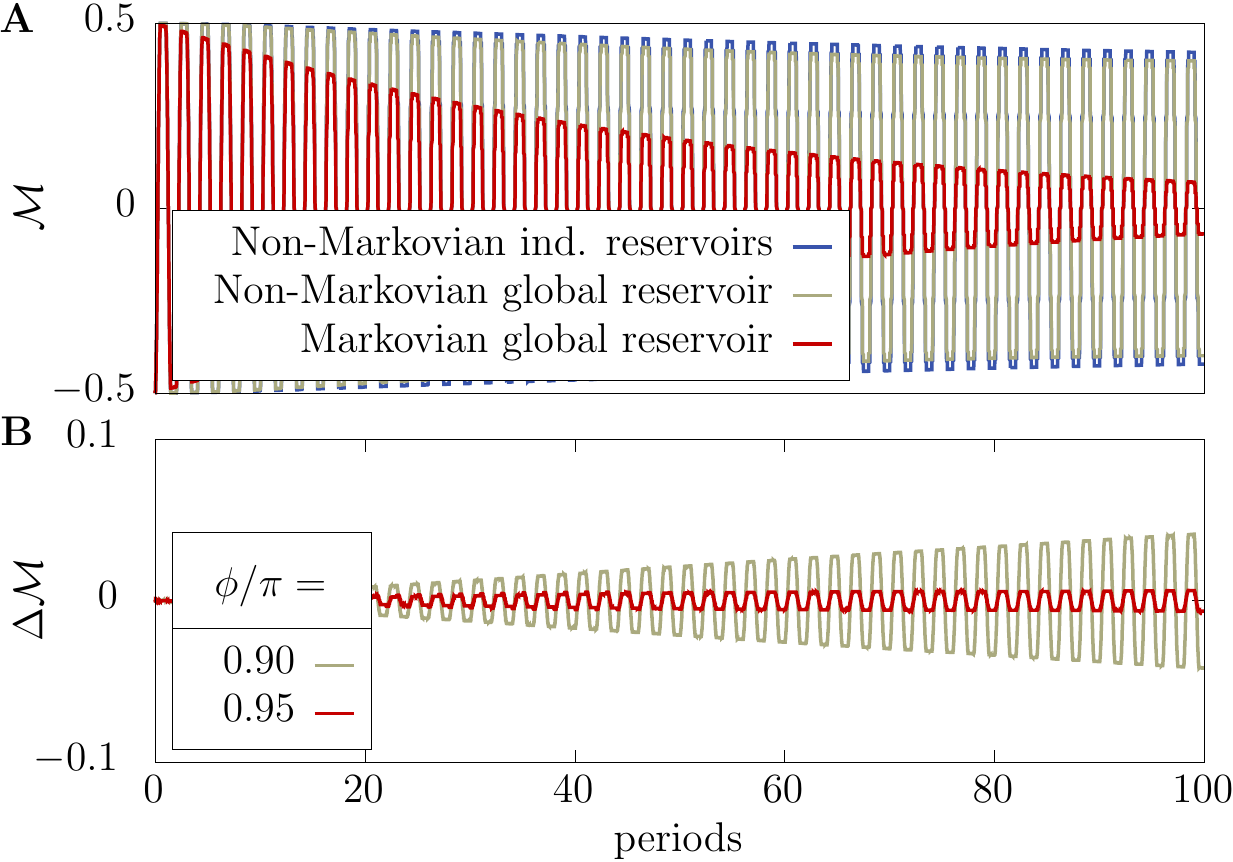}
\caption{(A) Comparison of the dynamics for different kinds of reservoirs ($\Gamma/J_z=1.0$) including an individual reservoirs for each of the spins (blue), a global reservoir (gray), and a Markovian reservoir (red) lacking the coherent feedback stabilization. (B) Robustness of the stabilization mechanism against deviations from the limit of optimal feedback phase.}
\label{fig:shared_ind}
\end{figure}
After having shown our main result that non-Markovian system-reservoir interaction leads
to a stabilization of the DTC, we discuss the robustness against imperfections
in the following.
In Fig.~\ref{fig:shared_ind}(A), we compare the non-Markovian case of individual reservoirs for each spin (blue) with the case of a global reservoir of all spins (gray).
Both, the global and the individual reservoir allow for a stabilization with only 
minor differences which saturates after $50$ periods.
Thus, the global reservoir creates initially more entanglement which is however subsequently suppressed, leading to a minor decrease in amplitude.
For comparison, the Markovian case for a global reservoir is included (red), leading quickly to equipartition as also for the individual reservoirs (not shown). 
While the stabilization works remarkably well in case of constructive feedback, it is important that fine-tuning of the feedback phases $\phi_{n_i}$ is not required.
In Fig.~\ref{fig:shared_ind}(B), we plot the deviations $\Delta \mathcal{M}$ in the dynamics of the staggered magnetization from the optimal case $\phi_{n_i}=\pi$ for different $\phi_{n_i}$.
For $\phi_{n_i}/\pi=0.95$, $\Delta \mathcal{M}$ saturates after around $40$ periods indicating that the feedback stabilization mechanism is robust against weak imperfections. For a larger $\phi_{n_i}/\pi=0.90$ the deviations from the stabilized signal still grow on the accessible time scales suggesting that strong deviations from the optimal feedback phase will eventually lead to a decay of the DTC signal.
\emph{Conclusion.}\textbf{--} We have shown that non-Markovian dynamics can stabilize the DTC, where conventional dissipation would result in thermalization with the environment. We demonstrated this stabilization for radiative decay in a kicked random Ising chain. Notice that time-crystalline signals in the presence of dissipation have been also reported in several other works recently~\cite{2015PhRvL.115p3601P,2015PhRvA..91c3617S,2018PhRvL.121c5301I,2019PhRvL.122a5701G}, where, however, the protecting nonergodicity avoiding equipartition is induced by integrability. This, however, does not represent a robust protecting mechanism in general since it can be lifted by infinitesimal perturbations~\cite{2016AdPhy..65..239D}. For the future it will be important to explore how our observations extend to other systems, other eigenstate phases such as the MBL spin glass, and other types of dissipation mechanisms such as dephasing.
\begin{acknowledgments}
We thank Andreas Knorr for helpful discussions. The authors gratefully acknowledge the support of the Deutsche Forschungsgemeinschaft (DFG)
through the project B1 of the SFB 910 and by the School of
Nanophotonics (SFB 787) and through the Gottfried Wilhelm Leibniz Prize program. The computations have been performed using the ITensor C++ library.
\end{acknowledgments}

%

\end{document}